\documentclass[nofootinbib,prl,aps,reprint,amsmath,amssymb]{revtex4-1}
\usepackage{hyperref}
\usepackage{changes}
\usepackage[section]{placeins}
\usepackage[titletoc]{appendix}
\usepackage{xcolor}
\usepackage{dsfont}
\usepackage{bm}
\usepackage{mathtools}
\usepackage{enumerate}
\DeclareMathAlphabet{\mathpzc}{OT1}{pzc}{m}{it}

\newcommand{\sayy}[1]{`#1'}
\providecommand{\href}[2]{#2}

\def\be{\begin{equation}}
\def\ee{\end{equation}}
\def\bea{\begin{eqnarray}}
\def\eea{\end{eqnarray}}

\def\sig{\sigma}

\def\la{\langle}
\def\ra{\rangle}
\def\Eu{ \mathfrak{H} }
\def\Sp{ \mathcal{S} }
\def\obs{\mathcal{O}}
\def\emi{\mathcal{E}}

\definecolor{MyB}{rgb}{0.1,0.1,1.0}

\begin{document}
\title{Redshift drift as a model independent probe of dark energy} 
\author{Asta~Heinesen}
\email{asta.heinesen@ens--lyon.fr}
\affiliation{Univ Lyon, Ens de Lyon, Univ Lyon1, CNRS, Centre de Recherche Astrophysique de Lyon UMR5574, F--69007, Lyon, France}

\begin{abstract}
It is well known that positive values of redshift drift is a signature of dark energy within the conventionally studied Friedmann-Lema\^{\i}tre-Robertson-Walker (FLRW) universe models. 
Here we show -- without making assumptions on the metric tensor of the Universe -- that redshift drift is a promising direct probe of violation of the strong energy condition within the theory of general relativity. 
\end{abstract}
\keywords{Redshift drift, relativistic cosmology, observational cosmology} 

\maketitle

\section{Introduction}
In general relativistic theory energy conditions are physically motivated constraints, which can be applied to the energy momentum tensor of space-time.
Energy conditions are important tools for constraining the possible solutions of the Einstein field equations and for deriving general theorems about the nature of gravitating systems. 
The Penrose and Hawking singularity theorems \cite{Penrose:1964wq,Hawking:1966sx} use energy conditions to arrive at physical scenarios where the developments of singularities are unavoidable. 
In particular, the strong energy condition is a central assumption in the focusing theorem, which states the conditions under which a matter congruence develops singularities in finite proper time.  

Within the Friedmann-Lema\^{\i}tre-Robertson-Walker (FLRW) framework of cosmology the strong energy condition is considered abandoned by observations \cite{Visser:1999de,Santos:2007pp}. The most direct evidence for violation of the strong energy condition, comes from the observed acceleration of space when interpreting data from supernovae of type Ia within the FLRW class of models \cite{acceleration_discovery_Riess,acceleration_discovery_Perlmutter,acceleration_perlmutter_1998}. 
However, no model independent falsification of the strong energy condition has been made to date in cosmology. 

Measurements of the drift of redshift in proper time of the observer \cite{Sandage,McVittie,Loeb:1998bu} are promising probes of the expansion history of the Universe. 
In the conventionally studied FLRW universe models, a positive value of redshift drift is a signature of dark energy \cite{Bolejko:2019vni}.  
The direct detection of the time-evolution of redshift -- expected within one to a few decades of observation time with facilities such as CODEX and the Square Kilometer Array (SKA) \cite{Balbi:2007fx,Liske:2008ph} -- allows for model independent determination of kinematic properties of the Universe. 
The potential for doing model independent analysis with redshift drift data must be accompanied by a theoretical understanding of redshift drift in generic universe models. 

So far most theoretical studies of redshift drift have been done within the FLRW class of models -- though see  \cite{Uzan:2008qp,LTB_zdrift,Mishra:2012vi,Balcerzak:2012bv,Fleury:2014rea,Jimenez:2019cll,Koksbang:2019glb,Koksbang:2020zej,Korzynski:2017nas,Heinesen:2020pms}. Local inhomogeneities and anisotropies in general contribute with accumulated effects along null rays and systematic effects from the position of the observer. Thus, local structure alter measurements of redshift drift \cite{Heinesen:2020pms} resulting in a violation of the signal expected from FLRW modelling. Such effects from local inhomogeneity are not \emph{a priori} expected to be subdominant, and the size of the effects must ultimately be determined by data. 
This in turn raises the question if redshift drift as a probe of dark energy -- or more generally the strong energy condition -- is valid in universe models that are not subject to the FLRW idealisation. 
In this paper we propose a model independent test of the strong energy condition by redshift drift measurements.

\vspace{3pt} 
\noindent
\underbar{Notation and conventions:}
Units are used in which $c=1$. Greek letters $\mu, \nu, \ldots$ label spacetime
indices in a general basis. Einstein notation is used such that repeated indices are summed over.  
The signature of the spacetime metric $g_{\mu \nu}$ is $(- + + +)$ and the connection $\nabla_\mu$ is the Levi-Civita connection. 
Round brackets $(\, )$ containing indices denote symmetrisation in the involved indices and square brackets $[\, ]$ denote anti-symmetrisation. 
Bold notation $\bm V$ for the basis-free representation of vectors $V^\mu$ is used occasionally. 

\section{Redshift drift in a general space-time}
\label{sec:zdrift}
In this section we review the expression for redshift drift for a generic space-time congruence of physical observers and emitters -- for details, see\footnote{For another interesting representation of redshift drift in general space-time models, see \cite{Korzynski:2017nas}.} \cite{Heinesen:2020pms}. 
We consider a general space-time congruence of observers and emitters (henceforth referred to as the \sayy{observer congruence}) with worldlines generated by the 4--velocity field $\bm u$ and parametrised by the proper time function $\tau$. 
The redshift drift of light rays generated by the 4-momentum field $\bm k$ and passing from an emitter placed at spacetime point $\emi$ to an observer situated at spacetime point $\obs$ can be written 
\bea
\label{redshiftdriftdec}
\frac{d z}{d \tau} \Bigr\rvert_{\obs} = (1+z) \Eu_\obs  - \Eu_\emi + \Sp_{\emi \rightarrow \obs}  \, ,
\eea  
with 
\bea
\label{ederivative1}
\hspace*{-0.3cm} \Sp_{\emi \rightarrow \obs}  =   E_\emi \! \! \int_{\lambda_\emi}^{\lambda_\obs} \! \! \! d \lambda \, \mathcal{I} \, , \qquad \mathcal{I} &\equiv& - k^\nu \nabla_\nu \! \! \left( \! \frac{e^\mu \nabla_\mu E}{E^2}  \! \right)  \, , 
\eea   
where the function $\lambda$ satisfies $k^\mu \nabla_\mu \lambda =1$, and is an affine parameter along each null line. 
The redshift $z$ and photon energy function $E$ associated with the light rays are given by 
\bea
\label{def:Eevolution}
z \equiv \frac{E_\emi}{E_\obs} - 1 \, , \qquad E \equiv - k^\mu u_\mu \, , 
\eea  
and the change of the photon energy $E$ along a given null ray is given by 
\bea
\label{def:Eevolution}
 \Eu \equiv - \frac{ k^{\mu}\nabla_{\mu} E }{E^2} =  \frac{1}{3}\theta  - e^\mu a_\mu + e^\mu e^\nu \sigma_{\mu \nu}   \, .
\eea  
The spatial unit vector $\bm e$ describes the spatial propagation direction of the null ray relative to an observer comoving with $\bm u$, and is defined by the decomposition 
\bea
\label{kdecomp}
k^\mu = E(u^\mu - e^\mu) \, .
\eea  
The function $\Eu$ is an observationally natural generalisation of the Hubble parameter of FLRW space-time: $\Eu_\obs$ plays the role of the proportionality constant between redshift and distance in the generalised Hubble law valid for arbitrary space-times in the $\mathcal{O}(z)$ vicinity of the observer \cite{Heinesen:2020bej}. The variables $\theta$, $ \sigma_{\mu \nu}$ and $a^\mu$ describe the expansion, shear and 4--acceleration of the observer congruence. Together with the vorticity tensor $\omega_{\mu \nu}$, they describe the kinematics of the observer congruence 
\bea
\label{def:expu}
&& \nabla_{\nu}u_\mu  = \frac{1}{3}\theta h_{\mu \nu }+\sig_{\mu \nu} + \omega_{\mu \nu} - u_\nu a_\mu  \ , \nonumber \\ 
&& \theta \equiv \nabla_{\mu}u^{\mu} \, ,  \quad \sig_{\mu \nu} \equiv h_{ \la \nu  }^{\, \beta}  h_{  \mu \ra }^{\, \alpha } \nabla_{ \beta }u_{\alpha  }  \, , \nonumber \\ 
&& \omega_{\mu \nu} \equiv h_{  \nu  }^{\, \beta}  h_{  \mu }^{\, \alpha }\nabla_{  [ \beta}u_{\alpha ] }   \, , \quad  a^\mu \equiv \dot{u}^\mu \,  , 
\eea 
where $h_{ \mu }^{\; \nu } \equiv u_{ \mu } u^{\nu } + g_{ \mu }^{\; \nu } $ is the spatial projection tensor defined in the frame of the 4--velocity field $\bm u$ and where triangular brackets $\la \ra$ denote traceless symmetrisation in the involved indices of a tensor in three dimensions\footnote{For two indices we have that the traceless parts of symmetric spatial tensors $T_{\mu \nu} = T_{(\mu  \nu)} = h_{ \mu }^{\, \alpha } h_{ \nu }^{\, \beta } T_{(\alpha  \beta)}$ is given by $T_{\la \mu \nu \ra}  = T_{\mu \nu} - \frac{1}{3}h_{\mu \nu} T$. Analogously for a tensor with three indices satisfying $T_{\mu \nu \rho} = T_{(\mu  \nu \rho)} = h_{ \mu }^{\, \alpha } h_{ \nu }^{\, \beta } h_{ \rho }^{\, \gamma } T_{(\alpha  \beta \gamma)}$, we have $T_{\la \mu \nu \rho \ra} = T_{\mu \nu \rho} - \frac{1}{5} \left( T_{\mu} h_{\nu \rho}  + T_{\nu} h_{\rho \mu} + T_{\rho} h_{\mu \nu} \right)$. For four indices we have  
$T_{\la \mu \nu \rho \kappa \ra}  = T_{\mu \nu \rho \kappa} - \frac{1}{7} ( T_{ \la \mu \nu \ra} h_{\rho \kappa} + T_{ \la \mu \rho \ra} h_{\nu \kappa}  + T_{ \la \mu \kappa \ra} h_{\nu \rho}  + T_{ \la \nu \rho \ra} h_{\mu \kappa}  
  +  T_{ \la \nu \kappa \ra} h_{ \mu \rho} + T_{ \la  \rho \kappa \ra} h_{  \mu \nu}   )  -  \frac{1}{5}T  h_{ ( \mu \nu}h_{\rho \kappa)} $. We have used the short hand notations $T_{ \mu \nu} \equiv h^{\rho \kappa} T_{ \mu \nu \rho \kappa}$, $T_{ \mu} \equiv  h^{\nu \rho } T_{ \mu \nu \rho}$, and $T \equiv h^{\mu \nu} T_{\mu \nu}$. 
}. 
The operator $\dot{} \equiv u^\mu \nabla_\mu$ denotes the derivative in proper time along flow lines of $\bm u$. 
From geometrical identities, the evolution of the kinematic variables $\theta$, $ \sigma_{\mu \nu}$ and $\omega_{\mu \nu}$ along the observer flow lines can be expressed as 
\bea
\label{raychaudhuri}
\hspace*{-0.3cm} \dot{\theta} 
& = & -\frac{1}{3}\,\theta^{2} - \sig_{\mu \nu} \sig^{\mu \nu} + \omega_{\mu \nu} \omega^{\mu \nu} 
  \nonumber \\   
\hspace*{-0.3cm}  && - u^{\mu}u^\nu R_{\mu \nu}  + D_\mu a^\mu  + a_\mu a^\mu \ , \\ 
\label{shear}
\hspace*{-0.3cm} \dot{\sig}_{\mu \nu} 
& = & - \frac{2}{3} \theta \sig_{\mu \nu}  - \sig_{\alpha \la \mu} \sig^{\alpha}_{\; \nu \ra}  +  \omega_{\alpha \la \mu} \omega^{\alpha}_{\; \nu \ra}  + 2 a^\alpha \sigma_{\alpha ( \nu} u_{\mu)}   \nonumber  \\ 
\hspace*{-0.3cm} && + D_{\la \mu} a_{\nu \ra} \! + \! a_{\la \mu} a_{\nu \ra}  \! - \! u^\rho u^\sigma  C_{\rho \mu \sigma \nu}  \! - \! \frac{1}{2} h^{\alpha}_{\, \la \mu} h^{\beta}_{\, \nu \ra}  R_{ \alpha \beta }    \, , \\
\label{rotation}
\hspace*{-0.3cm} \dot{\omega}_{\mu \nu} 
& = & - \frac{2}{3} \theta \omega_{\mu \nu}  + 2 \sigma^{\alpha}_{\, [\mu} \omega_{\nu ] \alpha}  - D_{[\mu} a_{\nu]}  - 2 a^\alpha \omega_{\alpha [ \mu} u_{\nu]}   \ , 
\eea 
where $R_{\mu \nu}$ is the Ricci tensor of the space-time and $C_{\mu \nu \rho \sigma}$ is the Weyl tensor. 
The operator $D_\mu$ is the covariant spatial derivative\footnote{The acting of $D_\mu$ on a tensor field $T_{\nu_1 , \nu_2, .. , \nu_n }^{\qquad \quad \; \gamma_1 , \gamma_2, .. , \gamma_m }$ is defined as: $D_{\mu} T_{\nu_1 , \nu_2, .. , \nu_n }^{\qquad \quad \; \gamma_1 , \gamma_2, .. , \gamma_m } \equiv  h_{ \nu_1 }^{\, \alpha_1 } h_{ \nu_2 }^{\, \alpha_2 } .. h_{ \nu_n }^{\, \alpha_n }    \,  h_{ \beta_1 }^{\, \gamma_1 } h_{ \beta_2 }^{\, \gamma_2 } .. h_{ \beta_m }^{\, \gamma_m }    \, h_{ \mu }^{\, \sigma } \nabla_\sigma  T_{\alpha_1 , \alpha_2, .. , \alpha_n }^{\qquad \quad \; \beta_1 , \beta_2, .. , \beta_m }$ .
} 
defined on the 3-dimensional space orthogonal to $\bm u$. 
The integrand in (\ref{ederivative1}) is conveniently expressed in terms of the (differentiated) kinematic variables of the observer congruence in the following series expansion \cite{Heinesen:2020pms} 
\bea
\label{integrandexp}
\hspace*{-0.65cm} \mathcal{I} \! = \! \mathcal{I}^{\it{o}} \! \!+\! e^\mu \mathcal{I}^{\bm{e}}_\mu   \!+\!   d^\mu \mathcal{I}^{\bm{d}}_\mu   \!+\!   e^\mu \! e^\nu \mathcal{I}^{\bm{ee}}_{\mu \nu}  \!+\!  e^\mu \! d^\nu \mathcal{I}^{\bm{ed}}_{\mu \nu}  \!+\!   e^\mu \! e^\nu \!e^\rho \mathcal{I}^{\bm{eee}}_{\mu \nu \rho} 
\eea  
with coefficients 
\bea
\label{coef}
&& \mathcal{I}^{\it{o}} \equiv - \frac{1}{3} (4 \omega^{\mu \nu} \omega_{\mu \nu}  + D_{\mu} a^{\mu} + a^{\mu} a_{\mu}  )  - d^\mu d_\mu   \, , \nonumber  \\   
&&  \mathcal{I}^{\bm{e}}_\mu  \equiv   \frac{1}{3} D_\mu \theta  + \! \frac{1}{3}  \theta a_\mu   \! + \!   \frac{2}{5}  D_{\nu} \sigma^{\nu }_{\, \mu}  +  \frac{2}{5}  a^{ \nu} \sigma_{\mu \nu}  \! - \! 2 a^\nu \omega_{\mu \nu}  \, ,  \nonumber  \\   
&& \mathcal{I}^{\bm{d}}_\mu \equiv - 2  a_\mu \, ,   \nonumber  \\  
&& \mathcal{I}^{\bm{ee}}_{\mu \nu} \equiv - \! \left( \! 4 \omega_{\alpha \mu}  \sigma^{\alpha}_{\; \nu} +  4 \omega_{\alpha \la \mu}  \omega^{\alpha}_{\; \nu \ra}  \! + \! D_{\la \nu} a_{\mu \ra} \! + \! a_{\la \mu} a_{\nu \ra } \! \right)    , \nonumber  \\   
&& \mathcal{I}^{\bm{ed}}_{\mu \nu} \equiv  4 (\sigma_{\mu \nu}  - \omega_{\mu \nu}   )  \, ,  \nonumber  \\   
&& \mathcal{I}^{\bm{eee}}_{\mu \nu \rho} \equiv   D_{\la \rho} \sigma_{\mu \nu \ra} +  a_{ \la \mu} \sigma_{\nu \rho \ra }  \, , 
\eea  
where $d^\mu \equiv h^{\mu}_{\; \nu} e^\alpha \nabla_\alpha e^\nu$ denotes the spatially projected \sayy{4--acceleration} of $\bm e$. The magnitude of $\bm d$ can be seen as a measure of the failure of $\bm e$ to define an axis of local rotational symmetry, and is thus a quantification of the local departure from isotropy \cite{hve1996}. In deriving (\ref{coef}), it has been assumed that the null congruence is irrotational, such that $\nabla_{[\alpha} k_{\nu]} = 0$.  
Note that all coefficients in (\ref{coef}) with more than one spacetime index are traceless. 

\section{Model independent test of the strong energy condition} 
For the purpose of examining the strong energy condition, it is convenient to rewrite the expression for redshift drift (\ref{redshiftdriftdec}), such that the Ricci curvature of the space-time appears explicitly in the formula. 
For this purpose the following identity 
\bea
\label{ederivative}
\hspace*{-0.3cm}  (1+z) \Eu_\obs  - \Eu_\emi  =   E_\emi \! \! \int_{\lambda_\emi}^{\lambda_\obs} \! \! \! d \lambda \, \mathcal{A} \, , \quad \; \mathcal{A} &\equiv& k^\nu \nabla_\nu \! \! \left( \! \frac{\Eu}{E}  \! \right)  \,  
\eea   
will be useful. Combining (\ref{redshiftdriftdec}), (\ref{ederivative1}) and (\ref{ederivative}) gives 
\bea
\label{redshiftdriftint}
\frac{d z}{d \tau} \Bigr\rvert_{\obs} = E_\emi \! \! \int_{\lambda_\emi}^{\lambda_\obs} \! \! d \lambda \,   \Pi  \, , \qquad \Pi \equiv \mathcal{I} + \mathcal{A}     \, . 
\eea  
In the FLRW limit, the integrand $\Pi$ reduces to the well known length scale acceleration \sayy{$\ddot{a}/a$}, where $a$ is the uniform FLRW scalefactor. 
The function $-\mathcal{A}/\Eu^2$ enters in the \sayy{Hubble law} for generic space-times \cite{Heinesen:2020bej} as an effective deceleration parameter, replacing the FLRW deceleration parameter in the series expansion of luminosity distance in redshift -- see \cite{Heinesen:2020bej} for details.   
In a similar spirit as for $\Eu$ and $\mathcal{I}$, the function $\mathcal{A}$ can be written as a truncated multipole expansion 
\bea
\label{Umulti}
\hspace*{-0.65cm} \mathcal{A} \! = \! \mathcal{A}^{\it{o}} \! \!+\! e^\mu \! \mathcal{A}^{\bm{e}}_\mu   \!+\!     e^\mu \! e^\nu \! \mathcal{A}^{\bm{ee}}_{\mu \nu}   \!+\!   e^\mu \! e^\nu \!e^\rho \! \mathcal{A}^{\bm{eee}}_{\mu \nu \rho}  \!+\!   e^\mu \! e^\nu \!e^\rho \! e^\kappa \! \mathcal{A}^{\bm{eeee}}_{\mu \nu \rho \kappa} 
\eea  
with coefficients 
\bea
\label{Ucoef}
&& \mathcal{A}^{\it{o}} \equiv - \frac{1}{3} u^\mu u^\nu R_{\mu \nu}  +  \frac{2}{3}  D_{   \mu} a^{\mu  }  - \frac{3}{5} \sigma^{\mu \nu} \sigma_{\mu \nu} + \frac{1}{3} \omega^{\mu \nu} \omega_{\mu \nu}    \, , \nonumber  \\   
&&  \mathcal{A}^{\bm{e}}_\mu \equiv  - \frac{2}{3} \theta a_\mu  + a^\nu \sigma_{\mu \nu}  + a^\nu \omega_{\mu \nu}    \nonumber  \\   
&& \qquad  \; \;  - \frac{1}{3} D_{\mu} \theta -  \frac{2}{5}   D_{  \nu} \sigma^{\nu }_{\;  \mu  }    - h^{\nu}_{\mu} \dot{a}_\nu     \, ,  \nonumber  \\   
&& \mathcal{A}^{\bm{ee}}_{\mu \nu} \equiv    3 a_{\la \mu}a_{\nu \ra }     - 2 \sigma_{\alpha  \mu} \omega^\alpha_{\; \nu }   - \frac{9}{7} \sigma_{\alpha \la \mu} \sigma^\alpha_{\; \nu \ra } +  \omega_{\alpha \la \mu} \omega^\alpha_{\; \nu \ra }  \nonumber \\  
&& \qquad  \; \;  +   2 D_{  \la \mu} a_{\nu \ra }   \! - \! u^\rho u^\sigma  C_{\rho \mu \sigma \nu}  \! - \! \frac{1}{2} h^{\alpha}_{\, \la \mu} h^{\beta}_{\, \nu \ra}  R_{ \alpha \beta }  \, , \nonumber  \\   
&& \mathcal{A}^{\bm{eee}}_{\mu \nu \rho} \equiv  -  D_{ \la \rho} \sigma_{\mu \nu  \ra }    -  5  a_{ \la \mu} \sigma_{\nu \rho \ra }   \, , \nonumber  \\   
&& \mathcal{A}^{\bm{eeee}}_{\mu \nu \rho \kappa}  \equiv  3 \sigma_{\la \mu \nu } \sigma_{\rho \kappa \ra}  \, ,
\eea  
where $R_{\mu \nu}$ is the Ricci curvature of the space-time, and where we have used the deviation equations (\ref{raychaudhuri}) and (\ref{shear}). In the derivation of (\ref{Umulti}) we have used the definition (\ref{def:Eevolution}) to write $\mathcal{A} = k^\mu \nabla_\mu (\Eu) / E + \Eu^2$ together with the identity 
\bea
\label{kderive}
\hspace*{-0.6cm} \frac{k^\nu \nabla_\nu e^\mu}{E} &=&  (e^\mu \! - \! u^\mu) \Eu   -   e^\nu \! \! \left(\! \frac{1}{3} \theta h^{\mu}_{\; \nu}  + \sigma^{ \mu}_{\; \nu} + \omega^{\mu}_{\; \nu} \! \right) +  a^\mu \,  .
\eea  
By combining the multipole coefficients in (\ref{coef}) and (\ref{Ucoef}) of the same order, we finally have that the integrand in (\ref{redshiftdriftint}) can be expressed as 
\bea
\label{Pimulti}
\hspace*{-0.65cm} \Pi &=& \Pi^{\it{o}}    +  e^\mu \Pi^{\bm{e}}_\mu  + d^\mu \Pi^{\bm{d}}_\mu   +       e^\mu   e^\nu \Pi^{\bm{ee}}_{\mu \nu} + e^\mu   d^\nu \Pi^{\bm{ed}}_{\mu \nu}     \nonumber  \\    
&& +  \, e^\mu   e^\nu  e^\rho \Pi^{\bm{eee}}_{\mu \nu \rho}   +    e^\mu   e^\nu  e^\rho   e^\kappa \Pi^{\bm{eeee}}_{\mu \nu \rho \kappa} \, 
\eea  
with coefficients 
\bea
\label{Picoef}
&& \Pi^{\it{o}} \equiv  - \frac{1}{3} u^\mu u^\nu R_{\mu \nu} + \frac{1}{3}D_{\mu} a^{\mu} -  \frac{1}{3} a^\mu a_\mu  \nonumber  \\    
&& \qquad \; \;    -  d^\mu d_\mu  - { \frac{3}{5}} \sigma^{\mu \nu} \sigma_{\mu \nu} - \omega^{\mu \nu} \omega_{\mu \nu}    \, , \nonumber  \\   
&&  \Pi^{\bm{e}}_\mu  \equiv   - \frac{1}{3}  \theta a_\mu    +  \frac{7}{5}  a_{ \nu} \sigma^{\nu }_{\, \mu} - a^\nu \omega_{\mu \nu} - h^{\nu}_{\mu} \dot{a}_\nu \, ,  \nonumber  \\   
&& \Pi^{\bm{d}}_\mu \equiv - 2  a_\mu \, ,   \nonumber  \\  
&& \Pi^{\bm{ee}}_{\mu \nu} \equiv   2  a_{\la \mu}a_{\nu \ra }       -  {\frac{9}{7}} \sigma_{\alpha \la \mu} \sigma^\alpha_{\; \nu \ra }  - 3 \omega_{\alpha \la \mu} \omega^\alpha_{\; \nu \ra }   - 6 \sigma_{\alpha  \mu} \omega^\alpha_{\; \nu }   \nonumber \\ 
&& \qquad  \; \;   + D_{  \la \mu} a_{\nu \ra }  -  u^\rho u^\sigma  C_{\rho \mu \sigma \nu}   -  \frac{1}{2} h^{\alpha}_{\, \la \mu} h^{\beta}_{\, \nu \ra}  R_{ \alpha \beta }  \,   , \nonumber  \\   
&& \Pi^{\bm{ed}}_{\mu \nu} \equiv  4 (\sigma_{\mu \nu}  - \omega_{\mu \nu}   )  \, ,  \nonumber  \\   
&& \Pi^{\bm{eee}}_{\mu \nu \rho} \equiv  - 4  a_{ \la \mu} \sigma_{\nu \rho \ra }  \, , \nonumber  \\   
&& \Pi^{\bm{eeee}}_{\mu \nu \rho \kappa} \equiv {3} \sigma_{\la \mu \nu } \sigma_{\rho \kappa \ra}  \, .
\eea  
The multipole coefficients (\ref{Picoef}) are given in terms of kinematic and dynamic variables associated with the observer congruence and the Ricci curvature tensor $R_{\mu \nu}$. 
{While the coefficients in (\ref{coef}) and (\ref{Ucoef}) contain spatial gradients of $\theta$ and $\sigma_{\mu \nu}$, these contributions cancel in (\ref{Picoef}), and the only spatial gradients that remain are of the 4--acceleration $a^\mu$.}
The anisotropic function $\Pi$ reduces to the FLRW scale factor acceleration \sayy{$\ddot{a}/a$} in the idealised isotropic and homogeneous limit.  
In the FLRW limit, the only non-zero contribution is the first term $- \frac{1}{3} u^\mu u^\nu R_{\mu \nu}$ of the monopole contribution $\Pi^{\it{o}}$ -- all of the remaining terms in (\ref{Picoef}) arise from the contributions of inhomogeneity and anisotropy along the null rays. 
These modifications of the FLRW law for redshift drift need not be small or cancel under the integral sign (\ref{redshiftdriftint}), and the measurements of redshift drift cannot a priori be expected to obey FLRW predictions in realistic universe models with structure. 

\vspace{5pt} 
\underbar{The dominant monopole approximation:}
Let us consider the case where the monopole term $\Pi^{\it{o}}$ is dominant in the integral expression for redshift drift (\ref{redshiftdriftint}), such that the contributions to the integral from the remaining terms in the series expansion (\ref{Pimulti}) are small compared to the contributions from $\Pi^{\it{o}}$. This corresponds to the physical assumption that the systematic alignment of $\bm e$ and $\bm d$ with the fluid variables such as shear and 4--acceleration is weak over the length scales of photon propagation.\footnote{The cancelation of spatially projected traceless combinations of fluid kinematic variables has been argued to be a realistic scenario in space-times where a notion of statistical homogeneity and isotropy is present, and where structure is slowly evolving relative to the timescale it takes for photons to pass an approximate homogeneity scale \cite{rae2009,rae2010}. 
This suggested cancelation has been shown to not hold true in general, exemplified by the systematic alignment of the propagation direction $\bm e$ of the null ray with the positive eigenvector of the shear tensor in a Tardis space-time \cite{Lavinto:2013exa}. The level of accuracy of the dominant monopole approximation $\int_{\lambda_\emi}^{\lambda_\obs} \! \! d \lambda \,  \Pi  \approx \int_{\lambda_\emi}^{\lambda_\obs} \! \! d \lambda \,  \Pi^{\it{o}}$ for light propagation over cosmological distances must be tested under various model assumptions.} 
In this scenario, we have that all spatial directions of photon propagation can be treated on equal footing at lowest order, with the leading order expression for redshift drift 
\bea
\label{redshiftdriftmon}
\frac{d z}{d \tau} \Bigr\rvert_{\obs} = E_\emi \! \! \int_{\lambda_\emi}^{\lambda_\obs} \! \! d \lambda \,  \Pi^{\it{o}}     \,  . 
\eea  
From (\ref{Picoef}) we see that the only potentially positive contributions to $\Pi^{\it{o}}$ are from the terms $- \frac{1}{3} u^\mu u^\nu R_{\mu \nu}$ and $\frac{1}{3} ( D_{\mu} a^{\mu} - a^\mu a_\mu)$. The other terms entering the expression for $\Pi^{\it{o}}$ are non-positive and in general contribute with accumulated negative contributions to the measured redshift drift. 
In general relativistic theory, negative values of $u^\mu u^\nu R_{\mu \nu}$ is equivalent to violation of the strong energy condition. 
For general relativistic space-times in which integrated values of $\Pi$ are dominated by the monopole contribution $\Pi^{\it{o}}$ for light propagation over cosmological distances, the only physical mechanisms that might result in the detection of positive values for redshift drift are thus (i) a special 4--acceleration profile of the space-time congruence of observers yielding integrated positive values of $D_{\mu} a^{\mu} - a^\mu a_\mu$ along the detected null rays.\footnote{In general relativistic perfect fluid cosmologies, the 4--acceleration field is given by $a_{\mu} = - D_\mu (p) / (\epsilon + p)$, where $\epsilon$ is the energy density and $p$ is the pressure associated with the perfect fluid description \cite{Buchert:2001sa} (see also the generalisation in \cite{Buchert:2019mvq} to arbitrary general relativistic space-times). Accumulated positive values of $D_{\mu} a^{\mu} - a^{\mu} a_{\mu} = - D^\mu D_\mu (p) / (\epsilon + p) + D^\mu (p) D_\mu  ( \epsilon ) / (\epsilon + p)^2$ along null rays might in this case occur for specific pressure and energy density profiles.}; (ii) violation of the strong energy condition. 
This realisation is the main result of this paper: \emph{A measured positive value of redshift drift indicates that the strong energy condition is violated.} A positive detection of redshift drift is in principle possible without such a violation, but it requires a 4--acceleration profile of the observer congruence giving systematic contributions along the null rays through its gradient. Alternatively, contributions from systematic alignment of the direction variables $\bm{e}$ and $\bm{d}$ with the dynamic fluid variables $\Pi^{\bm{e}}_\mu$, $\Pi^{\bm{d}}_\mu$, $\Pi^{\bm{ee}}_{\mu \nu}$, etc., in (\ref{Picoef}) over the length scales of light-propagation can cause inaccuracy of the approximation (\ref{redshiftdriftmon}). 

In the monopole approximation (\ref{redshiftdriftmon}), inhomogeneities tend to act with negative contributions to the redshift drift signal. 
We might thus in general expect the redshift drift signal to be negative in the absence of sources violating the strong energy condition -- even for space-times exhibiting globally defined acceleration of large scale cosmological volume sections due to the \sayy{backreaction}\footnote{For overviews of backreaction in cosmological modelling, see, e.g., \cite{ellisbuchert,buchertrasanen}.} of cosmic structures. 
This expectation is consistent with the numerical findings in \cite{Koksbang:2019glb,Koksbang:2020zej} for general relativistic inhomogeneous models without a cosmological constant, where the only example of positive redshift drift signals was obtained in an unphysical space-time scenario with a source of negative energy density violating the strong energy condition \cite{Koksbang:2020zej}.

\section{Conclusion} 
\label{sec:conclusion} 
In FLRW universe models a positive detection of redshift drift implies a non-zero cosmological constant \cite{Sandage,McVittie,Loeb:1998bu} and hence violation of the strong energy condition. 
In this paper we have considered the redshift drift signal in a general space-time, and written it in a form useful for examining the sign of redshift drift and its link to the strong energy condition in general relativistic universe models with no symmetry assumptions imposed on the metric.  
The redshift drift signal can be written in terms of a physically interpretable multipole series, where the coefficients are given in terms of kinematic variables and 4--acceleration of the observer congruence along with Ricci and Weyl curvature variables. The monopole contribution in this series represents the isotropic contribution common for all directions on the sky of the observer congruence. 
In a Universe where this monopole contribution is statistically dominant in the integral over the light path, and where the 4--acceleration profile of observers is not of a special form, a measured positive value of redshift drift is a direct signature of violation of the strong energy condition.

\vspace{6pt} 
\begin{acknowledgments}
This work is part of a project that has received funding from the European Research Council (ERC) under the European Union's Horizon 2020 research and innovation programme (grant agreement ERC advanced grant 740021--ARTHUS, PI: Thomas Buchert). I thank Thomas Buchert for his reading of the manuscript, and Sofie Marie Koksbang for comments. 
\end{acknowledgments}


\end{document}